\documentclass[preprint,NumberedRefs]{JASAnew}

\usepackage{graphicx}                                      
\usepackage{dcolumn}                                       
\usepackage{bm}                                                
\usepackage{amsmath,amsfonts}                         
\usepackage{amsthm}
\usepackage{latexsym}                                         
\usepackage{lineno}                                              
\usepackage{array}                                             
\usepackage{adjustbox}                                       
\usepackage{ulem}                                               
\usepackage{xcolor}                                             
\usepackage{hyperref}                                         
\usepackage{balance}
\usepackage{IEEEtrantools}

\newcommand{\RN}[1]{%
	\textup{\uppercase\expandafter{\romannumeral#1}}%
}

\newcommand\vcent[1]{\vcenter{\hbox{#1}}}
\newcommand\loudspeaker[1][3]{\ensuremath{\vcent{\rule{.6ex}{.6ex}}\kern-.5ex%
        \vcent{\scalebox{.6}[1]{\rotatebox[origin=center]{90}{$\blacktriangle$}}}%
        \ifnum#1>0\relax\kern.05ex\vcent{\scalebox{.4}{\ttfamily)}}%
        \ifnum#1>1\relax\kern-.4ex\vcent{\scalebox{.56}{\ttfamily)}}%
        \ifnum#1>2\relax\kern-.55ex\vcent{\scalebox{.7}{\ttfamily)}}%
        \fi\fi\fi}%
}

\newtheorem{theorem}{Theorem}

\begin{document}
\title[Sound field separation]{Real-time separation of non-stationary sound fields on spheres}
\author{Fei Ma}                                                            
\email{Author to whom correspondence should be addressed. 
   Electronic mail: fei.ma@ieee.org}
\affiliation{College of Engineering and Computer Science,%
	The Australian National University, Canberra, Australian Capital Territory 0200, Australia}
\author{Wen Zhang}                                                  
\affiliation{Center of Intelligent Acoustics and Immersive Communications, Northwestern 
	Polytechnical University, Xi’an 710072, Shaanxi, China}
\affiliation{College of Engineering and Computer Science,%
	The Australian National University, Canberra, Australian Capital Territory 0200, Australia}
\author{Thushara D. Abhayapala}                            
\affiliation{College of Engineering and Computer Science,%
	The Australian National University, Canberra, Australian Capital Territory 0200, Australia}

\date{\today}

\begin{abstract}
The sound field separation methods can separate the target field from the 
interfering noises, facilitating the study of the acoustic characteristics  of the 
target source, which is placed in a noisy environment. 
However, most of the existing sound field separation methods are derived in the  
frequency-domain, thus are best suited for separating stationary 
sound fields. 
In this paper, a time-domain sound field separation method is developed 
that can separate the non-stationary sound field generated by the target 
source over a sphere in real-time. 
A spherical array sets up a boundary between the target source and the 
interfering sources, such that the outgoing field on the array is only 
generated by the target source. 
The proposed method decomposes the pressure and the radial particle 
velocity measured by the array into spherical harmonics coefficients, 
and recoveries the target outgoing field based on the time-domain
relationship between the decomposition coefficients and the 
theoretically derived spatial filter responses. 
Simulations show the proposed method can separate non-stationary 
sound fields both in free field and room environments, and over a longer  
duration with small errors.
The proposed method could serve as a foundation for developing future 
time-domain spatial sound field manipulation algorithms. 
\end{abstract}


\maketitle

\section{\label{sec:1}INTRODUCTION}
To investigate the working condition of a machine or the acoustic  characteristics
of a musical instrument, we can use a sensor array to measure the target sound 
field generated by them and analyze the sensor array measurement. 
However, in practical acoustic environments, such as inside rooms, there are 
disturbing sources, which produce disturbing sound fields. 
The disturbing sound fields and the room reflected sound fields will interfere 
with the target sound field and contaminate the sensor array measurements, 
making it difficult to study the target source characteristics. 
Nonetheless, the target source does not necessarily co-locates with the 
disturbing sources. By exploiting the spatial characteristics of the 
sound sources, it is possible to separate the target sound field from 
the interfering sound fields using a sensor array. Thus we can study 
the target source  through analyzing the separated target 
sound field. 

Over the years, a number of sound field separation (SFS) methods have been 
reported in the literature. The spherical wave expansion based method uses 
the spherical harmonic modes as the basic modeling functions of a sound field, 
and are capable of separating the outgoing and incoming fields on a spherical 
sensor array.~\cite{Williams1999-gk,Weinreich1980-wq,Melon2010-cn,Braikia2013-ny}
The Spatial Fourier Transform  based method uses the two-dimensional 
Fourier Transform to decompose the sound field into plane-wave components, 
and can separates the incident and reflected sound fields on a planar 
sensor array.~\cite{Tamura1990-pj}    
The statistically optimized near field holography method is also capable 
of performing SFS on a planar sensor array, and can mitigate the spatial 
window leakage problem of the Spatial Fourier Transform based method.~\cite{Jacobsen2007-ky} 
The recently developed boundary element based method~\cite{Langrenne2007-hz,
Langrenne2009-mf} and the equivalent source based method~\cite{bi2008sound,
Bi2012-aj,fernand2012} extend the application of SFS to arbitrary shape 
sensor arrays. Further, these two methods can separate the scattering 
from the target source surface and recover the free-field radiation of 
the target source.

The above mentioned SFS methods can separate the sound fields incoming 
from two sides of a sensor array (or two sensor arrays) apart. However, 
they are all derived in the frequency-domain. That is, they first accumulate 
and transfer a frame of the time-domain pressure (or particle velocity) 
measurements into the time-frequency domain using the Short Time Fourier Transform. 
They then conduct the SFS  at each time-frequency bin. 
The Short Time Fourier Transform process inevitably introduces the 
spectrum leakage problem and frame-delays into the SFS methods.~\cite{Oppenheim1997-tu} 
The frame-delay is acceptable if the target sound field is stationary 
(or quasi-stationary). 
Nonetheless, when dealing with fast changing non-stationary sound field, the 
frame-delay will make the SFS methods unable to recover the target field 
in real-time. That further makes the separated target field unable to be used in
time-critical applications, such as active noise control,~\cite{ancref} 
real-time beamforming,~and machine anomaly diagnosis.~\cite{Langrenne2007-hz}  
A time-domain SFS method, on the other head, can track the changes of
the target sound field without introducing the frame latency issue. 

Manipulations of the sound fields over spatial regions in the time-domain  
are difficult, and time-domain SFS methods are seldom developed. The only 
existing time-domain SFS methods are developed by Bi and his 
coauthors.~\cite{zhang2012separation,Bi2014-ot,Bi2016-yy,Bi2016-jd}
Based on the Spatial Fourier Transform, the method they developed can 
separate the sound field coming from two sides of a planar sensor array.~\cite{zhang2012separation,Bi2014-ot,Bi2016-jd}
Based on the interpolated time-domain equivalent source method,~\cite{Bi2016-yy} 
they further developed a method that can separate the sound generated by 
a particular source in a multiple non-stationary sources scene.
However, these time-domain methods are best suited for SFS 
in free field (or semi-anechoic rooms).
Because first, in room environments, due to wall reflections the interfering sounds
and the target sound can arrive at a planar sensor array in the same directions,~\cite{Kuttruff2016-oe,
allen1979image} such that the target sound and the interfering noise become indistinguishable.
Second, the equivalent source based method requires perfect knowledge 
of the location and geometry information of all the sources. 
However, in a room environment, the image source effect exists, and it is 
difficult to characterize the image source especially for an arbitrary 
shape room with irregular boundaries.~\cite{Kuttruff2016-oe,allen1979image}

In this paper, based on the spherical wave expansion, we develop a 
time-domain SFS method that can separate the non-stationary outgoing 
and incoming sound fields on a sphere in both free 
field and room environments. 
We first decompose the pressure and the radial particle velocity 
into corresponding spherical harmonics coefficients. 
The time-domain convolution between the spherical harmonic coefficients and 
the derived impulse response functions results in
the outgoing/incoming field on the array. 
The performance of the proposed method is confirmed by simulations, and compared 
with the Spatial Fourier Transform based method.
Possible applications of the proposed method include 
(1) reference signal generation for active noise control system,~\cite{ancref} 
(2) real-time beamforming,
(3) de-reverberation for speech recognition and sound field reproduction system, 
(4) machine working condition monitoring in a noisy environment.

This paper is organized as follows. We introduce the problem formulation and 
the frequency-domain SFS method based on spherical harmonic expansion in Sec. II. 
In Sec. III, we provide the theoretical derivation of the time-domain SFS 
method. Sec. IV introduces the practical implementations of the proposed method, 
whose effectiveness is validated by simulations in Sec. V, and Sec. VI concludes 
this paper.  

\section{System model}
\subsection{Problem formulation}	
\begin{figure*}[t]
\centering
\begin{minipage}[b]{0.45\linewidth}
\centering
\centerline{\includegraphics[width=6cm]{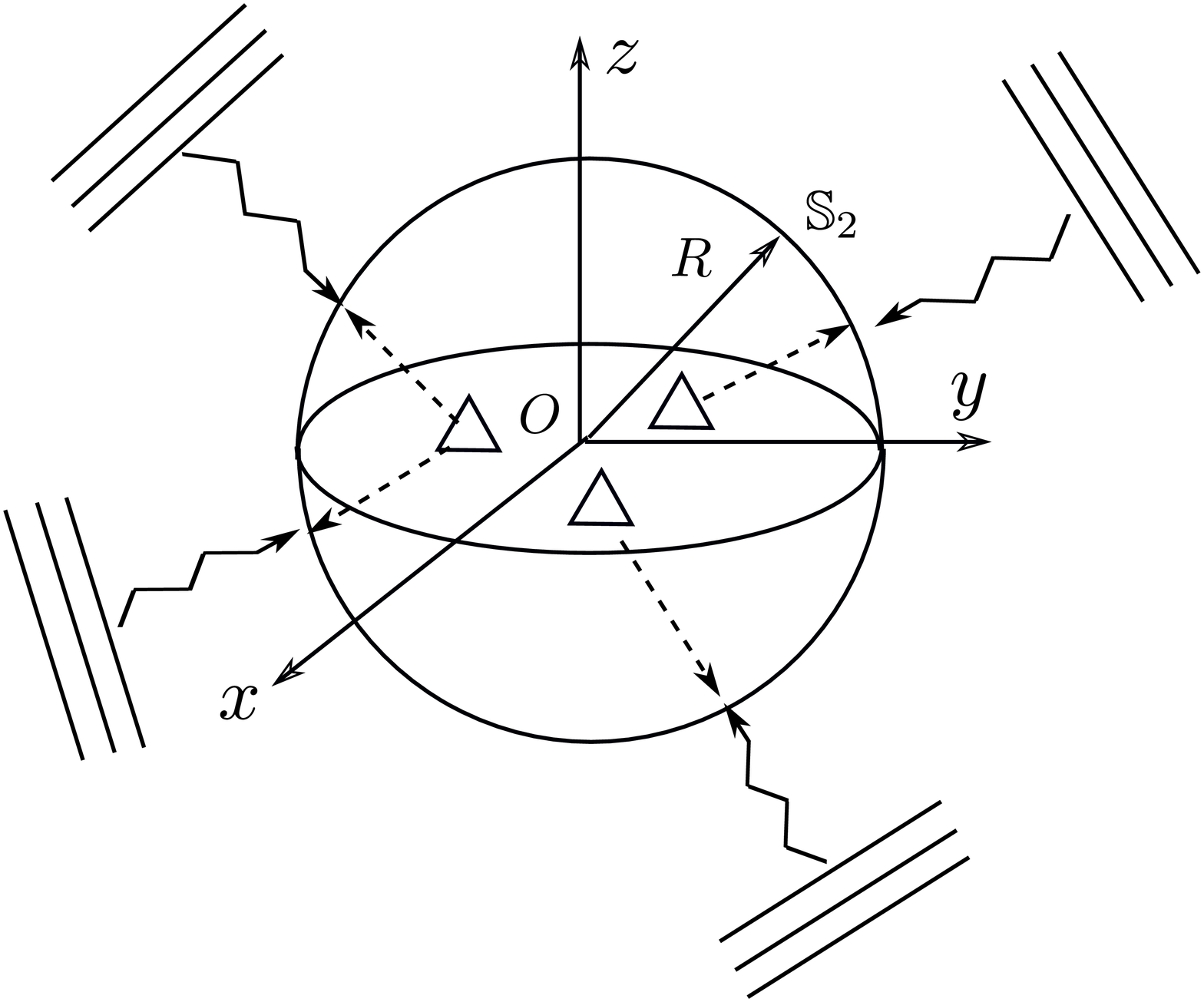}}
\centerline{(a)}
\end{minipage}
\centering
\begin{minipage}[b]{0.45\linewidth}
\centering
\centerline{\includegraphics[width=5cm]{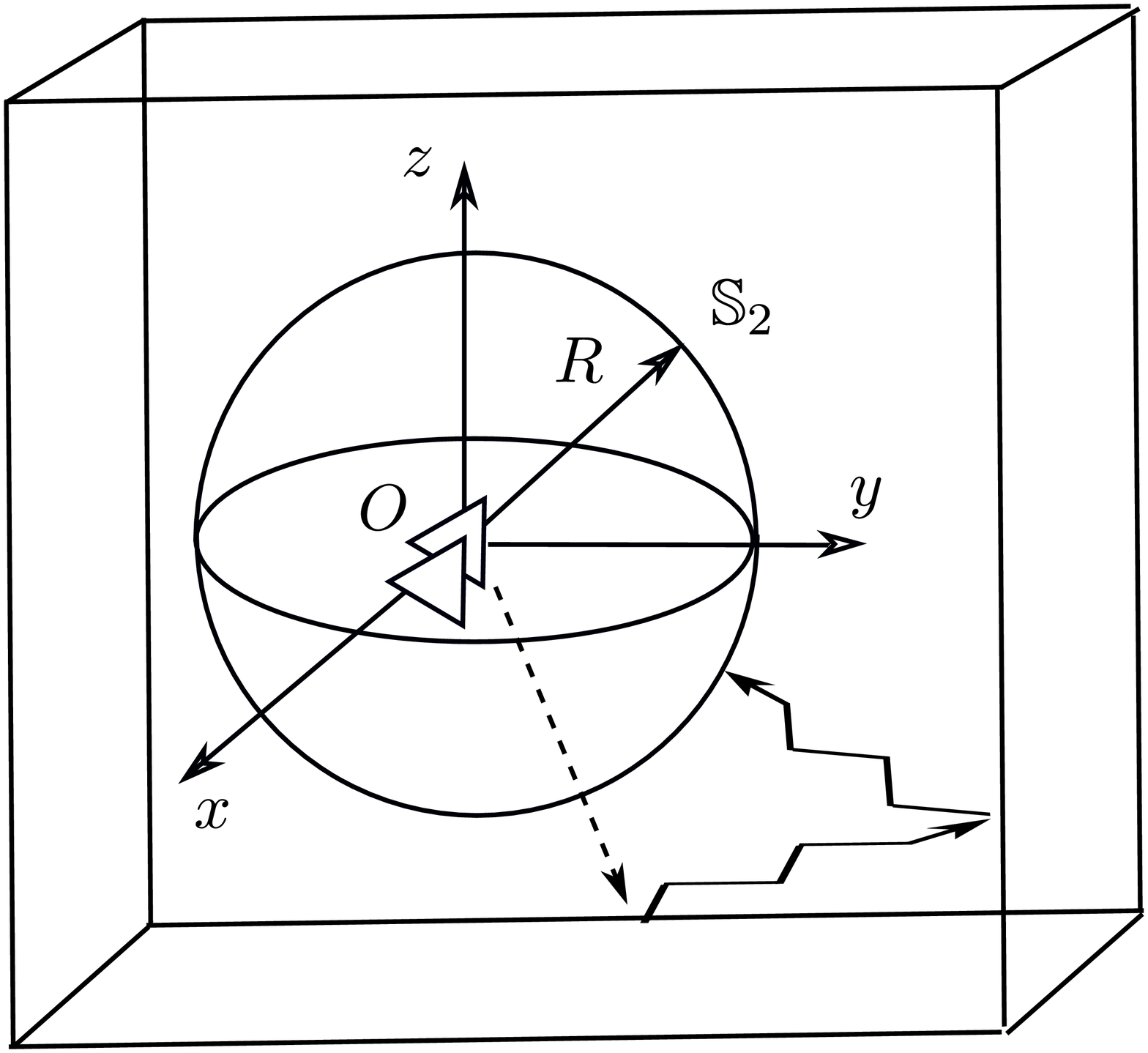}}
\centerline{(b)}
\end{minipage}
\caption{System models: (a) the target sources $\triangle$ are placed in free field, 
(b) the target sources $\triangle$  are placed inside a room. The outgoing and incoming 
field on the sphere $\mathbb{S}_2$ of radius $R$ are denoted as $\dashrightarrow$ 
and $\rightsquigarrow$, respectively.}
\label{fig:problem}
\end{figure*}

Consider the systems shown in Fig.~1, where in Fig.~1 
(a) the target sources (denoted as $\triangle$) are placed in free field, and in Fig.~1
(b) the target sources (denoted as $\triangle$) are placed inside a room.
We use $(x,y,z)$ and $(r,\theta,\phi)$ to denote the Cartesian and spherical coordinates 
of a point with respect to the point $O$, respectively.  
The sound field $p(t,R,\theta,\phi)$ on the sphere $\mathbb{S}_2$ of radius $R$ is the 
superposition of the outgoing field $p^{\mathrm{o}}(t,R,\theta,\phi)$ and the incoming field 
$p^{\mathrm{i}}(t,R,\theta,\phi)$
\begin{IEEEeqnarray}{rCl}
	p(t,R,\theta,\phi)&=&p^{\mathrm{o}}(t,R,\theta,\phi)+p^{\mathrm{i}}(t,R,\theta,\phi),
	\label{eq:pressure_sum1}
\end{IEEEeqnarray}
where $t$ is the continuous time.
The frequency-domain representation of~\eqref{eq:pressure_sum1} is 
\begin{IEEEeqnarray}{rCl}		
	\label{eq:pressure_sum22}
	P(\omega,R,\theta,\phi)
	&=& P^{\mathrm{o}}(\omega,R,\theta,\phi)+P^{\mathrm{i}}(\omega,R,\theta,\phi),	
\end{IEEEeqnarray}
where $\omega=2\pi{f}$ is the angular frequency ($f$ is the frequency).

As shown in Fig.~1, the incoming field $p^{\mathrm{i}}(t,R,\theta,\phi)$ 
on the sphere $\mathbb{S}_2$ is due to the external sources or room reflections. 
The outgoing field $p^{\mathrm{o}}(t,R,\theta,\phi)$ on the sphere $\mathbb{S}_2$, 
on the other hand, is generated by the target sources, thus characterizes the 
target sources. 
In this paper, we aim to separate the outgoing field $p^{\mathrm{o}}(t,R,\theta,\phi)$ 
on the sphere $\mathbb{S}_2$ to facilitate the study of the target sources. 

\subsection{\label{sec:frequency}The frequency-domain sound field separation on a sphere}

In the frequency-domain, the sound field $P(\omega,R,\theta,\phi)$ on a sphere $\mathbb{S}_2$ 
of radius $R$ consists of the outgoing field $P^{\mathrm{o}}(\omega,R,\theta,\phi)$ and the 
incoming field $P^{\mathrm{i}}(\omega,R,\theta,\phi)$~\cite{Williams1999-gk}
\begin{IEEEeqnarray}{rCl}		
	\label{eq:pressure_sum2}
	P(\omega,R,\theta,\phi)
	&=& P^{\mathrm{o}}(\omega,R,\theta,\phi)+P^{\mathrm{i}}(\omega,R,\theta,\phi)					\nonumber \\	
	&=&\sum_{\mu=0}^{\infty}\sum_{\nu=-\mu}^{\mu}A_{\mu\nu}(\omega,R)Y_{\mu\nu}(\theta,\phi),	    \nonumber \\
	&=&
	\sum_{\mu=0}^{\infty}\sum_{\nu=-\mu}^{\mu}
	[\underbrace{L_{\mu\nu}(\omega)h_{\mu}(\omega{}R/c)}
	_{A_{\mu\nu}^{\mathrm{o}}(\omega,R)} 
	+\underbrace{K_{\mu\nu}(\omega)j_{\mu}(\omega{R/c})}
	_{A_{\mu\nu}^{\mathrm{i}}(\omega,R)}]
	Y_{\mu\nu}(\theta,\phi),
\end{IEEEeqnarray}
where 
$c$ is the speed of sound, $A_{\mu\nu}(\omega,R)$ are spherical harmonic coefficients 
correspond the sound field $P(\omega,R,\theta,\phi)$ measured on the sphere $\mathbb{S}_2$, 
$A_{\mu\nu}^{\mathrm{o}}(\omega,R)$ and $A_{\mu\nu}^{\mathrm{i}}(\omega,R)$ are the 
outgoing and incoming field coefficients, respectively. $h_{\mu}(\cdot)$ is the second 
kind spherical Hankel function of order $u$, $j_{\mu}(\cdot)$ is the first kind spherical 
Bessel function of order $u$, $L_{\mu\nu}(\cdot)$ and $K_{\mu\nu}(\cdot)$ are the (radial-independent) 
spherical Hankel and Bessel function coefficients, respectively, 
%
$Y_{\mu\nu}(\cdot,\cdot)$ is the real spherical harmonic of order $n$ and degree $m$,~\cite{Poletti2009-zm} 
i.e.,
\begin{IEEEeqnarray}{rCl}
	\label{eq:rsh}
	Y_{\mu\nu}(\theta,\phi)\equiv
	&&\sqrt{\frac{(2\mu+1)(\mu-|\nu|)!}{4\pi(\mu+|\nu|)!}}\mathcal{P}_{\mu|\nu|}(\cos(\theta)) 
	\nonumber\\&&\times 	
	\begin{cases}
		\sqrt{2}\cos(|\nu|\phi),	& \nu>0,\\
		1,			 		   		& \nu=0,\\
		\sqrt{2}
		\sin(|\nu|\phi),			& \nu<0,\\
	\end{cases} 
\end{IEEEeqnarray}  
$\mathcal{P}_{\mu|\nu|}(\cdot)$ is the associated Legendre function of order $\mu$ and degree $|\nu|$.
Hereafter, we abbreviate $(\theta,\phi)$ as a single symbol $\Theta$ to simplify the notations. 

The frequency-domain radial particle velocity on the sphere $\mathbb{S}_2$ 
can be similarly expressed as~\cite{Williams1999-gk}
\begin{IEEEeqnarray}{rCl}  		
	\label{eq:velocity_sum2}
	V( \omega,R,\Theta)
	&=&\frac{i}{\rho_0{\omega}}\frac{\partial{}P(\omega,R,\Theta)}{\partial{r}}\Big|_{r=R} \nonumber\\ 
	&=&\sum_{\mu=0}^{\infty}\sum_{\nu=-\mu}^{\mu}B_{\mu\nu}(\omega,R)Y_{\mu\nu}(\Theta) \nonumber \\
	&=& \frac{i}{\rho_0{}c}
	\sum_{\mu=0}^{\infty}\sum_{\nu=-\mu}^{\mu}
	[L_{\mu\nu}(\omega)h_{\mu}^{\prime}(\omega{}R/c)                
	+K_{\mu\nu}(\omega)j_{\mu}^{\prime}(\omega{}R/c)]Y_{\mu\nu}(\Theta),
\end{IEEEeqnarray}
where $B_{\mu\nu}(\omega,R)$ are spherical harmonic coefficients corresponding to the 
radial particle velocity $V(\omega,R,\Theta)$ measured on the sphere $\mathbb{S}_2$, 
$h_{\mu}^{\prime}(\cdot)$ and $j_{\mu}^{\prime}(\cdot)$ are derivatives of $h_{\mu}(\cdot)$ 
and $j_{\mu}(\cdot)$ about the argument, respectively, $i$ is the unit imaginary number, 
and $\rho_0$ is the density of air.

Based on \eqref{eq:pressure_sum2} and \eqref{eq:velocity_sum2}, we obtain the outgoing 
and incoming field coefficients as~\cite{Williams1999-gk}
\begin{IEEEeqnarray}{rCl}
	\label{eq:f_out_coeff}
	{A}_{\mu\nu}^{\mathrm{o}}(\omega,R)
	&=&-i(\omega{}R/c)^2j_{\mu}^{\prime}(\omega{}R/c)h_{\mu} (\omega{}R/c) A_{\mu\nu}(\omega,R)      
	\nonumber\\
	&&+\rho_0{c}(\omega{}R/c)^2j_{\mu}(\omega{}R/c)h_{\mu} (\omega{}R/c){B}_{\mu\nu}(\omega,R),
\end{IEEEeqnarray}
\begin{IEEEeqnarray}{rCl}
	{A}_{\mu\nu}^{\mathrm{i}}(\omega,R)
	&=&
	i(\omega{R}/c)^2j_{\mu}(\omega{R/c})
	h_{\mu}^{\prime}(\omega{}R/c)	A_{\mu\nu}(\omega,R)	\nonumber\\
	&&-\rho_0c(\omega{}R/c)^2j_{\mu}(\omega{R/c})h_{\mu}(\omega{}R/c)B_{\mu\nu}(\omega,R),
\end{IEEEeqnarray}
respectively. 
Substitution of the coefficients  ${A}_{\mu\nu}^{\mathrm{o}}(\omega,R)$ and 
${A}_{\mu\nu}^{\mathrm{i}}(\omega,R)$ back into~\eqref{eq:pressure_sum2} 
results in the frequency-domain outgoing filed $P^{\mathrm{o}}(\omega,R,\Theta)$ 
and the incoming field  $P^{\mathrm{i}}(\omega,R,\Theta)$ on the sphere $\mathbb{S}_2$, 
respectively.



The effectiveness of the frequency-domain SFS method based on the spherical 
wave expansion has been validated by both simulations and experiments.~\cite{Melon2010-cn,Braikia2013-ny,Ma2018-ru}

The outgoing field separated by the frequency-domain method reveals the characteristics of the target source.
However, the frequency-domain method is based on the Short Time Fourier Transform 
to transfer the time-domain acoustic quantities into the time-frequency domain. 
The inherent frame-delay of the Short Time Fourier Transform process makes the 
separated outgoing field not suitable to use in  time-critical applications, 
such as active noise control and real-time beamforming.~\cite{ancref} 
A time-domain method, which does not introduce the frame-delay, is more 
appropriate to separate the non-stationary outgoing field for applications which 
have strict low-latency requirements.

\section{\label{sec:time}The time-domain sound field separation on a sphere}
In this section, we derive the time-domain SFS method on a sphere. We start
with a theorem, which provides the expressions of the outgoing/incoming field 
on a sphere. The detailed derivation of the theorem is in the appendix.

\begin{theorem}
	The time-domain outgoing field $p^{\mathrm{o}}(t,R,\Theta)$  and the incoming field 
	$p^{\mathrm{i}}(t,R,\Theta)$ on a sphere $\mathbb{S}_2$ of radius $R$ are 
	\begin{IEEEeqnarray}{rCl}
		p^{\mathrm{o}}(t,R,\Theta)=\sum_{\mu=0}^{\infty}\sum_{\nu=-\mu}^{\mu}
		a_{\mu\nu}^{\mathrm{o}}(t,R)Y_{\mu\nu}(\Theta),
		\label{eq:t_out}										
	\end{IEEEeqnarray}
	\begin{IEEEeqnarray}{rCl}
		p^{\mathrm{i}}(t,R,\Theta)=\sum_{\mu=0}^{\infty}\sum_{\nu=-\mu}^{\mu}
		a_{\mu\nu}^{\mathrm{i}}(t,R)Y_{\mu\nu}(\Theta),		\label{eq:t_inc}
	\end{IEEEeqnarray}
	where the outgoing field coefficients $a_{\mu\nu}^{\mathrm{o}}(t,R)$ and the incoming 
	field coefficients $a_{\mu\nu}^{\mathrm{i}}(t,R)$ are given by
	\begin{IEEEeqnarray}{rCl} 
		a_{\mu\nu}^{\mathrm{o}}(t,R)&=&\int_{\Theta}\Big\{
		\int_{0}^{2\tau_R}g_0^{\mu}(\tau)p(t-\tau,R,\Theta)
		d\tau	
		\nonumber\\
		&&+
		\int_{0}^{2\tau_R}g_1^{\mu}(\tau)\frac{d{}p(t-\tau,R,\Theta)}{dt}
	d\tau	
		\nonumber	 \\
		&&
		+\rho_0{}c
		\int_{0}^{2\tau_R}g_2^{\mu}(\tau)\frac{d{}v(t-\tau,R,\Theta)}{dt}d\tau
		\Big\}\nonumber\\ &&
				Y_{\mu\nu}(\Theta)d\Theta,			 
		\label{eq:t_out_coff}	
		\IEEEeqnarraynumspace
	\end{IEEEeqnarray}
	\begin{IEEEeqnarray}{rCl}
	a_{\mu\nu}^{\mathrm{i}}(t,R)&=&\int_{\Theta}\Big\{
		\int_{0}^{2\tau_R}g_3^{\mu}(\tau)p(t-\tau,R,\Theta)
		d\tau	
		\nonumber\\
		&&+
		\int_{0}^{2\tau_R}g_4^{\mu}(\tau)\frac{d{}p(t-\tau,R,\Theta)}{dt}
		d\tau	
		\nonumber \\
		&&
		-\rho_0{}c 
		\int_{0}^{2\tau_R}g_2^{\mu}(\tau)\frac{d{}v(t-\tau,R,\Theta)}{dt}d\tau\Big\}\nonumber\\
	&&	Y_{\mu\nu}(\Theta)d\Theta ,	\label{eq:t_inc_coff}
		\IEEEeqnarraynumspace
	\end{IEEEeqnarray}
	respectively, $\tau_R=R/c$, $p(t,R,\Theta)$ and $v(t,R,\Theta)$ are the pressure and the radial 
	particle velocity on the sphere at the point $(R,\Theta)$, respectively, and 
	$\int_{\Theta}d\Theta\equiv\int_{0}^{2\pi}\int_{0}^{\pi}\sin{}\theta{}d\theta{d}\phi$. 
	The expressions of $g_0^{\mu}(t)$, $g_1^{\mu}(t)$, $g_2^{\mu}(t)$, $g_3^{\mu}(t)$, 
	and $g_4^{\mu}(t)$ are
	\begin{IEEEeqnarray}{rCl}
		g_0^{\mu}(t)
		&=&\frac{\mu}{4\tau_R}\sum_{\nu=0}^{\mu} \sum_{\varsigma=0}^{\mu}\frac{\varphi_{\nu}(\mu)\varphi_{\varsigma}(\mu)}{(\nu+\varsigma)!}
		\Big[ (-1)^{\nu}(\frac{t}{\tau_R})^{\nu+\varsigma}\mathtt{Sign}(t)									
		\nonumber	\\
		&&
		+(-1)^{\mu+1} (\frac{t-2\tau_R}{\tau_R})^{\nu+\varsigma}\mathtt{Sign}(t-2\tau_R)		\Big]	,
		\label{eq:g0u}	
	\end{IEEEeqnarray}
	\begin{IEEEeqnarray}{rCl}
		g_1^{\mu}(t)
		&=& \frac{1}{4} \sum_{\nu=0}^{\mu+1}\sum_{\varsigma=0}^{\mu}\frac{\varphi_{\nu}(\mu+1)\varphi_{\varsigma}(\mu)}{(\nu+\varsigma)!}
		\Big[(-1)^{\nu}(\frac{t}{\tau_R})^{\nu+\varsigma}\mathtt{Sign}(t)									
		\nonumber	\\
		&&
		+(-1)^{\mu} (\frac{t-2\tau_R}{\tau_R})^{\nu+\varsigma} \mathtt{Sign}(t-2\tau_R)\Big],
		\label{eq:g1u}	
	\end{IEEEeqnarray}
	\begin{IEEEeqnarray}{rCl}
		g_2^{\mu}(t)
		&=&\frac{1}{4}\sum_{\nu=0}^{\mu} \sum_{\varsigma=0}^{\mu}\frac{\varphi_{\nu}(\mu)\varphi_{\varsigma}(\mu)}{(\nu+\varsigma)!}\Big[(-1)^{\nu}
		(\frac{t}{\tau_R})^{\nu+\varsigma}\mathtt{Sign}(t)		
		\nonumber	\\
		&&
		+ (-1)^{\mu+1}(\frac{t-2\tau_R}{\tau_R})^{\nu+\varsigma}\mathtt{Sign}(t-2\tau_R)\Big], 
		\label{eq:g2u}
	\end{IEEEeqnarray}
	\begin{IEEEeqnarray}{rCl}
		g_3^{\mu}(t)
		&=&\frac{\mu+1}{4\tau_R}\sum_{\nu=0}^{\mu} \sum_{\varsigma=0}^{\mu}\frac{\varphi_{\nu}(\mu)\varphi_{\varsigma}(\mu)}{(\nu+\varsigma)!}
		\Big[ (-1)^{\nu}(\frac{t}{\tau_R})^{\nu+\varsigma}\mathtt{Sign}(t)									
		\nonumber	\\
		&&
		+(-1)^{\mu+1} (\frac{t-2\tau_R}{\tau_R})^{\nu+\varsigma}\mathtt{Sign}(t-2\tau_R)		\Big]	,	
	\end{IEEEeqnarray}
	\begin{IEEEeqnarray}{rCl}
&&g_4^{\mu}(t)=
\begin{cases}
\frac{1}{4}\sum_{\nu=0}^{\mu}\sum_{\varsigma=0}^{\mu-1}\frac{\varphi_{\nu}(\mu)\varphi_{\varsigma}(\mu-1)}
{(\nu+\varsigma)!}\Big[(-1)^{\nu}(\frac{t}{\tau_R})^{\nu+\varsigma}\mathrm{Sign}(t)\nonumber\\
	\qquad+(-1)^{\mu-1}(\frac{t-2\tau_R}{\tau_R})^{\nu+\varsigma}\mathrm{Sign}(t-2\tau_R)\Big], \quad \mu>0\\
\frac{1}{4}\mathrm{Sign}(t)-\frac{1}{4}\mathrm{Sign}(t-2\tau_R),\quad\qquad\qquad\qquad\mu=0,
		\end{cases}	\\
	\end{IEEEeqnarray}
	respectively, where $\mathtt{Sign}(\cdot)$ is the sign function, 
	$\varphi_{\nu}(\mu)={(\mu+\nu)!}/({2^{\nu}\nu!(\mu-\nu)!})$ for 
	$\nu=0, 1, ..., \mu$, $\mu\geq0$.~\cite{Abramowitz1965-oq}
	Note that $g_0^{\mu}(t)={\mu}g_2^{\mu}(t)/{\tau_R}$ and 
	$g_3^{\mu}(\tau_R)={(\mu+1)}g_2^{\mu}(t)/{\tau_R}$.
\end{theorem}

We have the following remarks on the time-domain SFS method:
\begin{enumerate}
	\item Equations~\eqref{eq:t_out},~\eqref{eq:t_inc}, \eqref{eq:t_out_coff}, 
	\eqref{eq:t_inc_coff} show that the outgoing (incoming) field 
	at a particular point $(R,\Theta)$ 
	can be described by the pressure, the pressure gradient, 
	and the radial partial velocity gradient observed over the sphere within a certain time frame.
	\item The spatial filter functions $g_0^{u}(t)$, $g_1^{u}(t)$, $g_2^{u}(t)$, 
	$g_3^{u}(t)$, and $g_4^{u}(t)$ are specially designed such that they take finite 
	values inside of the time interval $[0,2\tau_R]$ and are zeros outside of the interval. 
	This limits the integrals~\eqref{eq:t_out_coff}~and~\eqref{eq:t_inc_coff} 
	to the interval $[0,2\tau_R]$.
	\item Figure~2 depicts ${g}_0^{\mu}(t)$, ${g}_1^{\mu}(t)$, $g_2^{\mu}(t)$ and 
	their estimations ${\hat{g}_0^{\mu}}(t)$, ${\hat{g}_1^{\mu}}(t)$, ${\hat{g}_2^{\mu}}(t)$  
	for $u=0,1,2$, $R=0.5$ m, $c=343$ m/s, and $\tau_R=R/c\approx1.458$ ms. We plot  
	${g}_0^{\mu}(t)$, ${g}_1^{\mu}(t)$, $g_2^{\mu}(t)$ based on the theoretical 
	expressions \eqref{eq:g0u}, \eqref{eq:g1u}, and \eqref{eq:g2u}, respectively.
We plot	${\hat{g}_0^{\mu}}(t)$, ${\hat{g}_1^{\mu}}(t)$, and ${\hat{g}_2^{\mu}}(t)$ 
based on the numerical calculation of~\eqref{eq:g0uexp},~\eqref{eq:g1uexp}, and~\eqref{eq:g2uexp}, 
	respectively. As shown in Fig.~2, the theoretically derived ${g}_0^{\mu}(t)$, 
	${g}_1^{\mu}(t)$, $g_2^{\mu}(t)$ agree with the numerically calculated ${\hat{g}_0^{\mu}}(t)$, 
	${\hat{g}_1^{\mu}}(t)$, and ${\hat{g}_2^{\mu}}(t)$, respectively. 
\begin{figure}[t]
\centering
\begin{minipage}[b]{0.9\linewidth}
\centering
\centerline{\includegraphics[trim={0.4cm 0 0.3cm 0},clip,width=8cm]{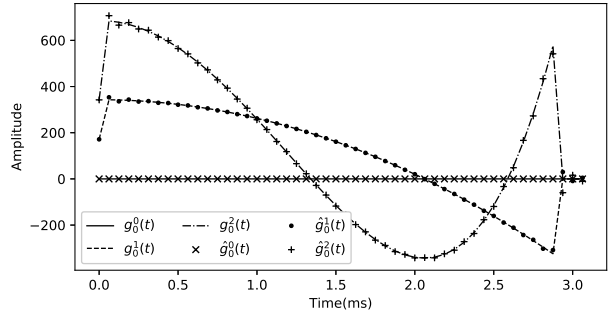}}
\centerline{(a)}\medskip
\end{minipage}
\centering
\begin{minipage}[b]{0.9\linewidth}
\centering
\centerline{\includegraphics[trim={0.4cm 0 0.3cm 0},clip,width=8cm]{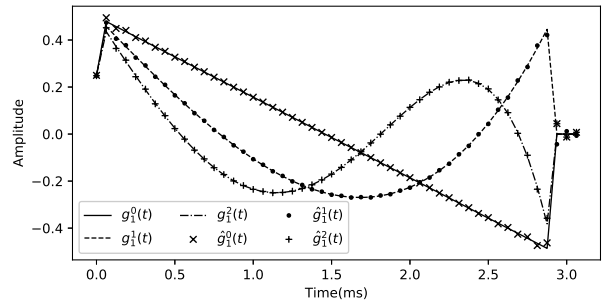}}
\centerline{(b)}\medskip
\end{minipage}
\centering
\begin{minipage}[b]{0.9\linewidth}
\centering
\centerline{\includegraphics[trim={0.4cm 0 0.3cm  0},clip,,width=8cm]{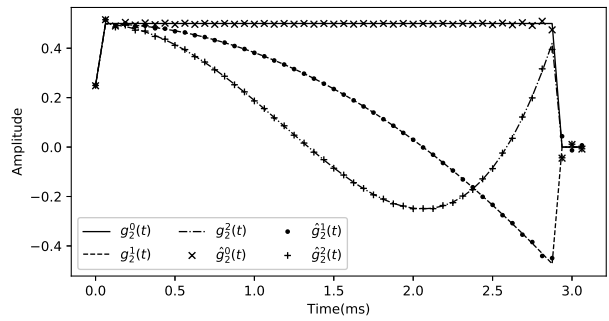}}
\centerline{(c)}\medskip
\end{minipage}
\caption{The plots of the theoretically derived $g_0^{\mu}(t)$, $g_1^{\mu}(t)$, 
$g_2^{\mu}(t)$ and the numerically calculated ${\hat{g}_0^{\mu}}(t)$, ${\hat{g}_1^{\mu}}(t)$, 
${\hat{g}_2^{\mu}}(t)$ for $\mu=0,1,2$, $R=0.5$ m, $c=343$ m/s, and $\tau_R\approx1.458$ ms.} 
\label{fig:gut}
\end{figure}
	\item The frequency-domain outgoing/incoming filed coefficients 
	$\{A_{\mu\nu}^{\mathrm{o}}(\omega,R),A_{\mu\nu}^{\mathrm{i}}(\omega,R) \}$ and their 
	time-domain counterparts $\{a_{\mu\nu}^{\mathrm{o}}(t,R),a_{\mu\nu}^{\mathrm{i}}(t,R)\}$ 
	depend on the radius $R$ of the sound field separation sphere $\mathbb{S}_2$. 
	We can use the radial-independent spherical Bessel and Hankel function coefficients 
	$\{L_{\mu\nu}(\omega),K_{\mu\nu}(\omega)\}$~in~\eqref{eq:pressure_sum2} to characterize 
	the outgoing/incoming field, and derive their corresponding time-domain counterparts~$\{l_{\mu\nu}(t),k_{\mu\nu}(t)\}$. 
	By manipulating $\{l_{\mu\nu}(t),k_{\mu\nu}(t)\}$, we not only can recover 
	the outgoing/incoming field on a sphere, but also are able to project the 
	outgoing/incoming field from one sphere to another.~\cite{Williams1999-gk} 
	This follow up work is out the scope of the paper and 
	 will be one of our future works. 
\end{enumerate}

\section{\label{subsec:realize}Realization of the time-domain sound field separation method}
The theoretically derived~\eqref{eq:t_out},~\eqref{eq:t_inc},~\eqref{eq:t_out_coff}, 
and~\eqref{eq:t_inc_coff} of the proposed method involve continuous integrals, 
time derivatives of the sound pressure and the radial particle velocity, which are 
acoustic quantities that can not be measured directly using existing sensors.
In this section, we further develop approaches to facilitate the implementation 
of the proposed  method. 

\subsection{Discretization and sampling }

We first approximate the continuous integral in \eqref{eq:t_out_coff} by a discrete form 
as follows  
\begin{IEEEeqnarray}{rCl} 
	a_{\mu\nu}^{\mathrm{o}}(t,R) 
	&
	{{\approx}}
	&
    \int_{\Theta}\Big\{	{\sum_{n=0}^{T_n}}
	g_0^{\mu}(\tau_n) {p(t-\tau_n,R,\Theta)}{\delta_\tau}	 	\nonumber	 \\
	&&+{\sum_{n=0}^{T_n}}
	g_1^{\mu}(\tau_n)\frac{p(t-\tau_n,R,\Theta)-p(t-\delta_t-\tau_n,R,\Theta)}{\delta_\tau}	 
	{\delta_t}\nonumber	 \\
	&&+\rho_0{c}{\sum_{n=0}^{T_n}}g_2^{\mu}(\tau_n)
	\frac{v(t-\tau_n,R,\Theta)-v(t-\delta_t-\tau_n,R,\Theta)}{\delta_t}{\delta_\tau}\Big\} \nonumber\\
    &&	Y_{\mu\nu}(\Theta)d\Theta \nonumber		 \\ 
	&\approx&\int_{\Theta}\Big\{
	\sum_{n=0}^{T_n}g_0^{\mu}(\tau_n)
	p(t-\tau_n,R,\Theta)\delta_t	\nonumber \\
	&&+\sum_{n=0}^{T_n}g_1^{\mu}(\tau_n)
	[p(t-\tau_n,R,\Theta)-p(t-\delta_t-\tau_n,R,\Theta)]		\nonumber \\
	&&+
	\rho_0{c}
	\sum_{n=0}^{T_n}g_2^{\mu}(\tau_n)
	[v(t-\tau_n,R,\Theta)-v(t-\delta_t-\tau_n,R,\Theta)]\Big\}\nonumber\\
	&&Y_{\mu\nu}(\Theta)d\Theta,	
	\label{eq:t_out_coff1}
\end{IEEEeqnarray}
where 
$\delta_\tau=\delta_t=1/f_\mathrm{s}$, and $T_n=\lceil{}2f_\mathrm{s}\tau_R\rceil+1$ 
is the number of  samples corresponding to $2\tau_R$ ($f_\mathrm{s}$ 
is the sampling frequency, and $\lceil\cdot\rceil$ is the ceiling operator).

We further develop~\eqref{eq:t_out_coff1} by replacing the continuous integral
over the sphere with a finite summation at  $(R,\Theta_q)_{q=1}^{Q}$ and by 
substituting the continuous time $t$ and $\tau_n$ using the discrete 
time $n$ and $n'$.
By denoting $p(n,R,\Theta_q)$ and $v(n,R,\Theta_q)$ as the pressure and the radial 
partial velocity at $(R,\Theta_q)$ at discrete time instant $n$, 
we have the outgoing field coefficient $a_{\mu\nu}^{\mathrm{o}}(n,R)$
at discrete time $n$ as
\begin{IEEEeqnarray}{rCl}
	a_{\mu\nu}^{\mathrm{o}}(n,R)
	&\approx& 
	\sum_{q=1}^{Q}\gamma_q Y_{\mu\nu}(\Theta_{q})\sum_{n'=0}^{T_n} 
	\Big\{g_0^{\mu}(n')p(n-n',R,\Theta_q)   {\delta_t} 	\nonumber\\
	&&+g_1^{\mu}(n') [p(n-n',R,\Theta_q) - p(n-1-n',R,\Theta_q)] 			 \nonumber\\
	&&+
	\rho_0{c}g_2^{\mu}(n')[v(n-n',R,\Theta_q)- v(n-1-n',R,\Theta_q)]  
	\Big\},
	\label{eq:discreterealization}
\end{IEEEeqnarray}
where $\{\gamma_q\}_{q=1}^{Q}$ are the sampling weights.~\cite{Rafaely2015-yh} 

Substitution of $a_{\mu\nu}^{\mathrm{o}}(t,R) $ by $a_{\mu\nu}^{\mathrm{o}}(n,R)$ 
into~\eqref{eq:t_out} results in an estimation of the outgoing field 
$p^{\mathrm{o}}(n,R,\Theta)$ at discrete time instance on the sphere $\mathbb{S}_2$. 
We can obtain the discrete form of~\eqref{eq:t_inc_coff} in a similar manner.  

Due to the finite sampling over the sphere, we can only recover 
the outgoing field coefficients up to $N$, where $N\leq\sqrt{Q}$ is the 
truncation order of the outgoing field.~\cite{ward2001reproduction}

\subsection{Pressure and velocity approximation}
We may use an array of acoustic vector sensors, which can measure both the pressure 
and the radial particle velocity,~\cite{De_Bree2003-eg} to realize the 
SFS method. 
Alternatively, we can use two microphone arrays to realize the method. 

Place $Q$ microphones on a sphere of radius 
$R-\delta_R$ at $\{R-\delta_R,\Theta_q\}_{q=1}^{Q}$, and another 
$Q$ microphones on a sphere of radius $R+\delta_R$ at 
$\{R+\delta_R,\Theta_q\}_{q=1}^{Q}$, where $\delta_R\ll{}R$ is 
a real positive number. We approximate the pressure 
on the middle sphere of radius $R$ at positions $\{R,\Theta_q\}_{q=1}^{Q}$ 
by~\cite{Parkins2000-ny}
\begin{IEEEeqnarray}{rCl}
	\label{eq:pressureapproximation}
	p(n,R,\Theta_q)\approx 0.5[p(n,R+\delta_R,\Theta_q)+p(n,R-\delta_R,\Theta_q)].
\end{IEEEeqnarray}
Based on the Euler's equation, we approximate the radial particle velocity on the 
middle sphere of radius $R$  at positions 
$\{R,\Theta_q\}_{q=1}^{Q}$ by~\cite{Parkins2000-ny} 
\begin{IEEEeqnarray}{rCl}
	\label{eq:velocityapproximation}
	v(n,R,\Theta_q)&\approx& v(n-1,R,\Theta_q) \nonumber\\
	&&+  \frac{1}{\rho_0{}}
	\frac{p(n,R+\delta_R,\Theta_q)-p(n,R-\delta_R,\Theta_q)}{2\delta_R}\delta_t.
\end{IEEEeqnarray}

Substitutions of~\eqref{eq:pressureapproximation} and~\eqref{eq:velocityapproximation} 
into~\eqref{eq:discreterealization} result in an estimation of the time-domain outgoing 
field coefficient $a_{\mu\nu}^{\mathrm{o}}(n,R)$.

\section{Simulation examples}
In this section, we conduct simulations to validate the 
effectiveness of the proposed time-domain SFS method.

\subsection{\label{sec:vs}Sound field separation in free field}
The simulation environment is as shown in Fig.~1 (a). We place 
a point source at $(0,0,0.3)$ m as the target source, and 100 plane waves, 
whose directions are determined according to the 100-point spherical 
packing,~\cite{packing} as the incoming fields. 
We let the target source output  a band-limited sound, which is simulated by 
passing a unit-variance Gauss white noise through a 64-tap Bandpass Butterworth 
filter of the frequency range $[100,600]$ Hz. 
The incoming plane waves have the same frequency components as the target source 
output. The strengths of the plane waves are normalized such that their contributions 
to the sound pressure on the sphere are approximately the same compared with the 
contribution from the target source.

We aim to estimate the target (outgoing) sound field on the sphere $\mathbb{S}_2$ 
produced by the target source.  We choose the radius of the sphere $\mathbb{S}_2$ 
as $R=0.65$ m, and the outgoing field truncation order to be 
$N=5$. We use a sensor array arranged on the sphere according to the 6-th order Gauss 
sampling scheme to realize the time-domain SFS method.~\cite{Rafaely2015-yh} 
We simulate the impulse responses, including the radial particle velocity 
response, between the source and the sensors based on the free-space Green 
function and the Euler's equation.~\cite{Williams1999-gk} The impulse responses 
are truncated to 1024 taps long under the sampling frequency $f_\mathrm{s}=48$ 
kHz. The speed of sound is $c=343$ m/s, and the air density is $\rho_0=1.225$ 
kg/m$^3$. The length of impulse response functions $g_0^{\mu}(t)$, $g_1^{\mu}(t)$, 
$g_2^{\mu}(t)$ is $T_n=\lceil{2R/cf_{\mathrm{s}}}\rceil+1=183$  taps.
We add Gauss white noise to 
the sensor measurement such that the signal to noise ratio is 40 dB. The 
simulation results are from an average of 100 independent runs. The settings 
in this paragraph are used in all simulations unless otherwise stated. 

We reconstruct the outgoing field on the sphere over a period of 10 ms 
according to~\eqref{eq:t_out}. We depict the amplitudes of the outgoing 
field $p^{\mathrm{o}}(t,R,\Theta_{17})$, the total field  $p(t,R,\Theta_{17})$, 
the separated outgoing field $\hat{p}^{\mathrm{o}}(t,R,\Theta_{17})$, 
and the outgoing field separation error $p^{\mathrm{e}}(t,R,\Theta_{17})=
p^{\mathrm{o}}(t,R,\Theta_{17})-\hat{p}^{\mathrm{o}}(t,R,\Theta_{17})$,
at the 17-th sensor in Fig.~3. 
The total filed is the summation of the outgoing field and the incoming 
plane wave fields.  
\begin{figure}[t]
\centering
\includegraphics[width=9cm]{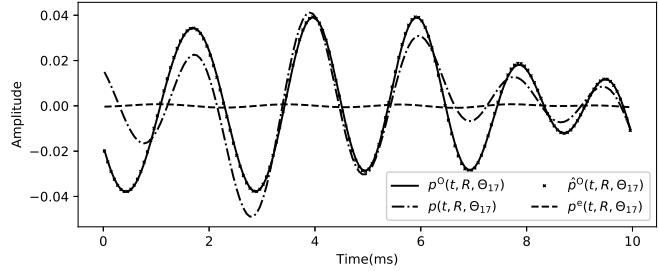}
\caption{
Sound field separation at a point: The amplitudes of the outgoing field 
$p^{\mathrm{o}}(t,R,\Theta_{17})$, the total field  $p(t,R,\Theta_{17})$, 
the separated outgoing field $\hat{p}^{\mathrm{o}}(t,R,\Theta_{17})$, and 
the outgoing field separation error $p^{\mathrm{e}}(t,R,\Theta_{17})$,
at the 17-th sensor over 10 ms.} 
\label{fig:gfree}
\end{figure}

As shown in Fig.~3, the total field differs from the outgoing 
field. The separated outgoing field approximates the outgoing field, and the 
field separation error is small over the whole 10 ms period. The normalized 
separation error at the 17-th sensor over the observation period is 
\begin{IEEEeqnarray}{rCl}
\xi_{17}=10\mathrm{log}10
\frac{\sum_{t=0}^{10\; \mathrm{ms}}||p^{\mathrm{e}}(t,R,\Theta_{17})||^2}
{\sum_{t=0}^{10\;\mathrm{ms}}||p^{\mathrm{o}}(t,R,\Theta_{17})||^2}=-30.1\; 
\mathrm{dB}.\IEEEeqnarraynumspace
\end{IEEEeqnarray}

We depict the amplitudes of the outgoing field $p^{\mathrm{o}}(t,R,\Theta_{q})$, 
the total field  $p(t,R,\Theta_{q})$, the separated outgoing field $\hat{p}^{\mathrm{o}}
(t,R,\Theta_{q})$, and the outgoing field separation error $p^{\mathrm{e}}(t,R,\Theta_{q})
=p^{\mathrm{o}}(t,R,\Theta_{q})-\hat{p}^{\mathrm{o}}(t,R,\Theta_{q})$, 
on the sphere $\mathbb{S}_2$ at $t=10$ ms  in Fig.~4. 
Here $(R,\Theta_q)_{q=1}^{180\times360}$ are equal-angle sampling point 
positions on the sphere $\mathbb{S}_2$.~\cite{Rafaely2015-yh} As shown 
in Fig.~4, the proposed method is able to recover 
the outgoing field accurately over the whole sphere, and the normalized
separation error over the whole sphere is 
\begin{IEEEeqnarray}{rCl}
	\xi_{\mathbb{S}_2}
	=10\mathrm{log}10\frac{\sum_{q=1}^{180\times360}||p^{\mathrm{e}}(t,R,\Theta_{q})||^2}
	{\sum_{q=0}^{180\times360}||p^{\mathrm{o}}(t,R,\Theta_{q})||^2}=-29.5\; \mathrm{dB}. 
	\IEEEeqnarraynumspace
\end{IEEEeqnarray}

\begin{figure*}[t]
\centering
\begin{minipage}[b]{0.475\linewidth}
\centering
\centerline{\includegraphics[trim={0cm 1cm 0cm 2cm},clip,width=8.5cm]{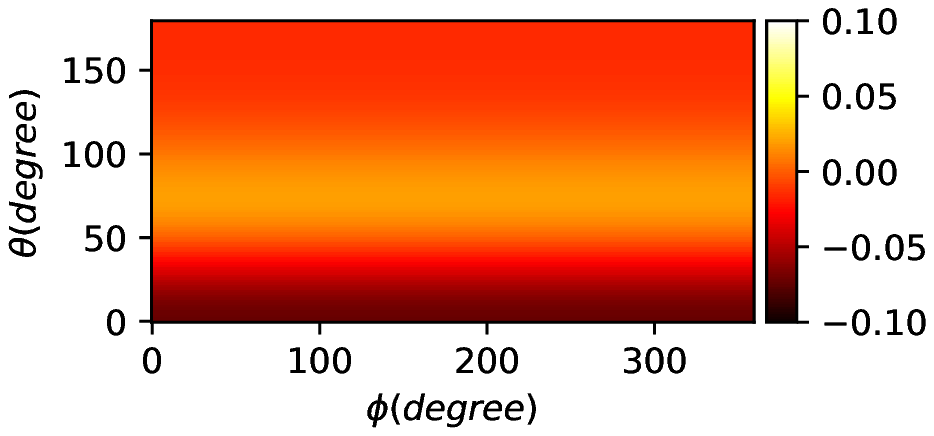}}
\centerline{(a)}
\end{minipage}
\centering
\begin{minipage}[b]{0.475\linewidth}
\centering
\centerline{\includegraphics[trim={0cm 1cm 0cm 2cm},clip,width=8.5cm]{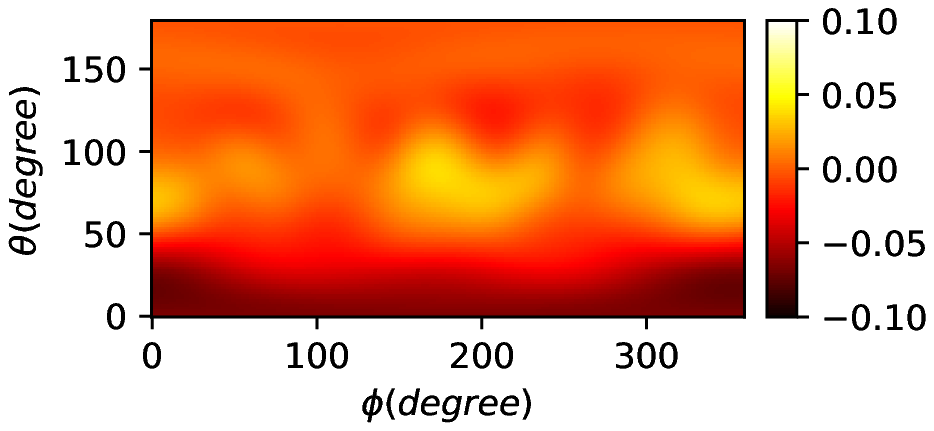}}
\centerline{(b)}
\end{minipage}
\centering
\begin{minipage}[b]{0.475\linewidth}
\centering
\centerline{\includegraphics[trim={0cm 1cm 0cm 2cm},clip,width=8.5cm]{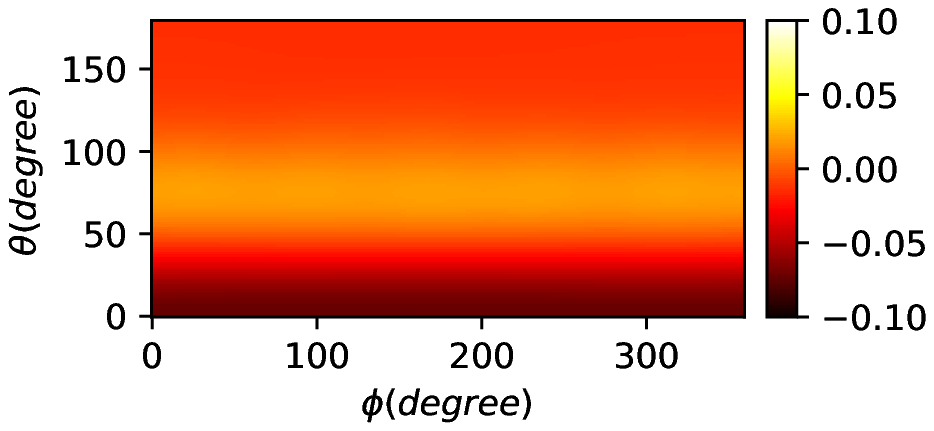}}
\centerline{(c)}
\end{minipage}
\centering
\begin{minipage}[b]{0.475\linewidth}
\centering
\centerline{\includegraphics[trim={0cm 1cm 0cm 2cm},clip,width=8.5cm]{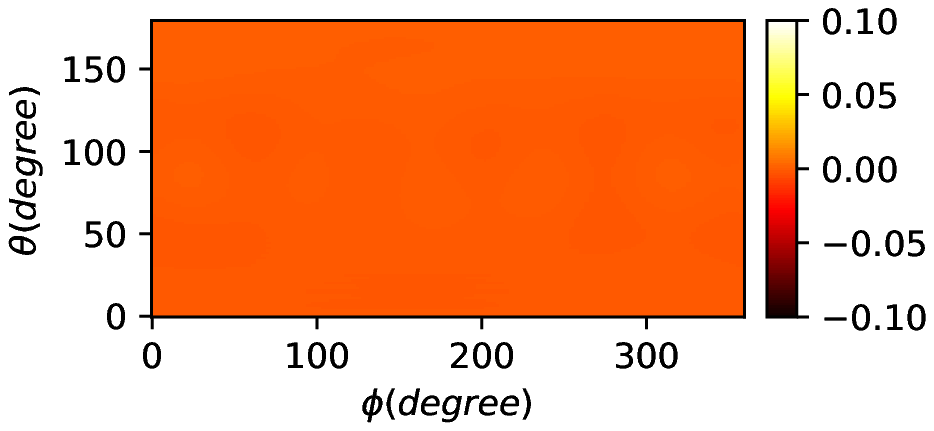}}
\centerline{(d)}
\end{minipage}
\caption{(Color online) Sound field separation over the whole sphere: The 
amplitudes of (a) the outgoing field $p^{\mathrm{o}}(t,R,\Theta_{q})$, (b) 
the total field $p(t,R,\Theta_{q})$, (c) the separated outgoing field 
$\hat{p}^{\mathrm{o}}(t,R,\Theta_{q})$, and (d) the outgoing field 
separation error $p^{\mathrm{e}}(t,R,\Theta_{q})$, on the sphere 
$\mathbb{S}_2$ at $t=10$ ms. Here $(R,\Theta_q)_{q=1}^{180\times360}$ are 
equal angle sampling point positions.~\cite{Rafaely2015-yh}}
\label{fig:gfreewhole}
\end{figure*}

We further investigate the performance of the proposed method in terms of the 
frequency and the truncation order $N$. Define the normalized separation error 
over all the sampling points as   
\begin{IEEEeqnarray}{rCl}
\xi^{N}(\omega)=
10\mathrm{log}10\frac{\sum_{q=1}^{98}||p^{\mathrm{o}}(\omega,R,\Theta_{q})-\hat{p}^{\mathrm{o},N}(\omega,R,\Theta_{q})||^2}
{\sum_{q=1}^{98}||p^{\mathrm{o}}(\omega,R,\Theta_{q})||^2},\nonumber\\
\end{IEEEeqnarray}
where $p^{\mathrm{o}}(\omega,R,\Theta_{q})$ is the frequency-domain outgoing field, 
$\hat{p}^{\mathrm{o},N}(\omega)$ is the frequency-domain counterpart of the following 
equation 
\begin{IEEEeqnarray}{rCl}
	\hat{p}^{\mathrm{o},N}(n,R,\Theta_q)=\sum_{\mu=0}^{N}\sum_{\nu=-\mu}^{\mu}
	\hat{a}_{\mu\nu}^{\mathrm{o}}(n,R)Y_{\mu\nu}(\Theta_q),
	\label{eq:t_out_e}										
\end{IEEEeqnarray}
$\hat{a}_{\mu\nu}^{\mathrm{o}}(n,R)$ are the time-domain outgoing field 
coefficient obtained through~\eqref{eq:discreterealization},
$\{R,\Theta_{q}\}_{q=1}^{98}$ are the sampling point positions from the 
6-th order Gauss sampling scheme, and $\omega=200\pi, 202\pi,..., 1200\pi$ 
rad/s (or $f=100, 101, ..., 600$ Hz).

\begin{figure}[t]
\centering
\includegraphics[width=8cm]{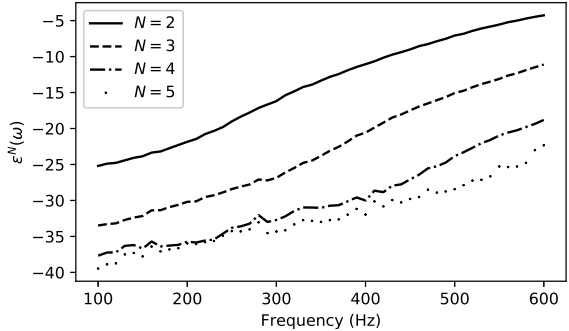}
\caption{separation error: The normalized separation error as a function of 
the	frequency and the truncation order $N$. 	
} 
\label{fig:frequency}
\end{figure}

We plot the normalized separation error $\xi^{N}(\omega)$ in Fig.~\ref{fig:frequency},
which shows that $\xi^{N}(\omega)$  increases along with the frequency. 
The higher the outgoing field truncation order $N$ is the smaller the normalized 
separation error $\xi^{N}(\omega)$ is. However, in the low frequency range, 
the truncation order $N=5$ does not make the normalized separation error 
$\xi^{N}(\omega)$ significantly smaller than that in the case of $N=4$, i.e., 
$\xi^{5}(\omega)\approx\xi^{4}(\omega)$ for $\omega<800$ $\pi$/rad or 
$f<400$ Hz. To conclude, we should choose an appropriate truncation order 
$N$ for the separated target (outgoing) field  depending on its frequency 
components such that the separation error is sufficiently small.     

\subsection{\label{sec:compare}Comparisons with the Spatial Fourier Transform based method }
In this section, we compare the performance of the proposed method 
with the Spatial Fourier Transform based  method.~\cite{Bi2014-ot}

The setup of the simulation environment is shown in Fig.~6, 
which has  a Cartesian coordinate system with the origin at the point $O$. 
On the $z=0$ m plane $H$, we equally distribute 15$\times$15 sensors on a 
square of size 0.7 m $\times$ 0.7 m, whose center is at the point $O$. 
On the $z=0.01$ m plane $H_1$, we place another 15$\times$15 sensors equally 
on a square of size 0.7 m $\times$ 0.7 m, whose center is the at the point 
$(x,y,z)=(0.0,0.0,0.01)$ m. We use the sensors on these two planes for SFS 
using the Spatial Fourier Transform based method. We extend the array size to be 81$\times$81 by 
zero-padding to reduce the aliasing error of the Spatial Fourier Transform.~\cite{Bi2014-ot} 

There is a sphere $\mathbb{S}_2$ of radius $R=0.35$ m. On the sphere 
$\mathbb{S}_2$, we place sensors according to the 9-th order Gauss sampling 
scheme.~\cite{Rafaely2015-yh} We use the sensors on this sphere 
$\mathbb{S}_2$ to estimate the outgoing field up to 6-th order using 
the proposed method. 

The sampling schemes used by these two methods are selected such 
that they can detect approximately the same number of acoustic 
quantities, and that the sizes of the two arrays are comparable. 

We have a target point source at $(0.0,0.0,-0.2)$ m, and another interfering 
point source at $(0,0,0.2)$ m as shown in Fig.~6. Outputs 
of these two sources are unit-variance Gauss white noises filtered a 64-tap 
Butterworth Bandpass filter of frequency range $[100, 1000]$ Hz. 
In the Spatial Fourier Transform based method, we truncate the Bessel 
function, i.e., (7) from Bi,~\cite{Bi2014-ot} to 1024 taps long. In the 
proposed method, the length of impulse response functions $g_0^{\mu}(t)$, 
$g_1^{\mu}(t)$, $g_2^{\mu}(t)$ is $T_n=\lceil{2R/cf_{\mathrm{s}}}\rceil+1=99$ 
taps. 

To make the comparison fair, only the outgoing sound at the point $O$ (the sound 
produced by the target source in the positive $z$ direction), where the plane 
$H$ is tangent to the sphere $\mathbb{S}_2$, is separated by both methods. 
The overall duration of the SFS process is about 30 ms.

\begin{figure}[t]
\centering
\includegraphics[width=6cm]{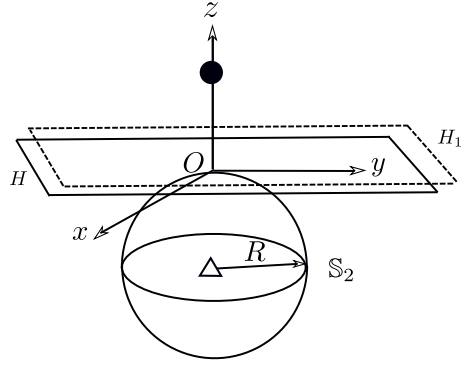}
\caption{Setup of sound field separation a point: The proposed method use 
the sensors on the sphere $\mathbb{S}_2$ to separate the outgoing field 
at the point $O$, while the Spatial Fourier Transform based method use 
sensors on plane $H$ and $H_1$ to separate the sound waves at the point $O$ 
produced by the target source $\triangle$ and the disturbing source $\bullet$.} 
\label{fig:comparewith}
\end{figure}

\begin{figure*}[t]
\centering
\includegraphics[width=18cm]{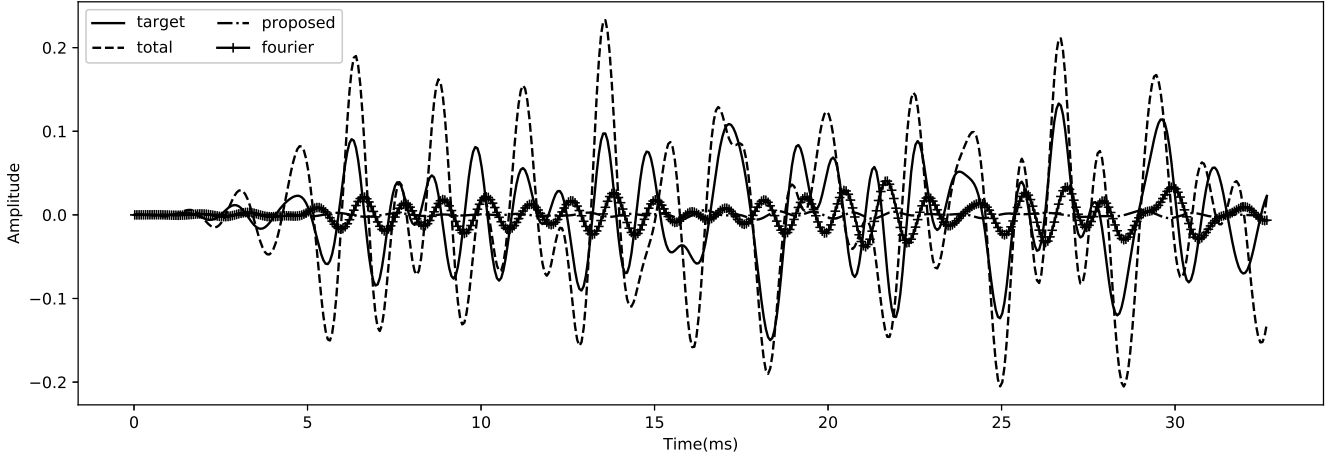}
\caption{Sound field separation at the point $O$: The amplitudes 
of the target sound, the total sound, the sound separation errors 
using the proposed and the Spatial Fourier Transform based method over 
50 ms.} 
\label{fig:compare}
\end{figure*}

The target (outgoing) sound $p^\mathrm{d}(t)$, the total sound $p^\mathrm{t}(t)$, 
and the sound separation error $p^\mathrm{e}(t)=p^\mathrm{d}(t)-\hat{p}^\mathrm{d}(t)$ 
using the proposed and the Spatial Fourier Transform-based method are shown in Fig.~7. 
The target (outgoing) sound is produced by convolving the target source output and 
the free-space Green function. The total sound is the summation of the target 
(outgoing) sound and the interfering (incoming) sound. 
The separated target sounds $\hat{p}^\mathrm{d}(t)$ in the two methods are not 
shown for the ease of illustration.

As shown in Fig.~7, the proposed method can recover the target 
sound accurately over the whole 30 ms period. The Spatial Fourier Transform 
based method can separate the target field with small errors before $t=5$ ms, 
but after that the SFS error starts to increase. That is because, in the 
Spatial Fourier Transform based method, the target sound at one time instant 
depends on all previous sounds.~\cite{Bi2014-ot} The truncation of the infinite 
long Bessel function, i.e., (7) from Bi,~\cite{Bi2014-ot} to a finite length 
introduces errors.
We have conducted simulations using the Spatial Fourier Transform based method 
with a longer truncation length (8192 taps) for the Bessel functions. 
The sound field separation error are similar to the dotted line as shown in 
Fig.~7, thus are not shown for brevity.
In the proposed method we have specially designed the spatial filter functions 
functions, $g_0^{\mu}(t)$, $g_1^{\mu}(t)$, $g_2^{\mu}(t)$, to be finite 
long, avoiding SFS error due to the impulse response function truncation. 

The simulation results demonstrate that the Spatial Fourier Transform-based 
method, which does not require the sensor array to surround the target source, 
is best suited for separating transient plane wave sound fields. The proposed 
method, on the other hand, can accurately separate the target (outgoing) sound 
field on a spherical array over a longer duration.

\subsection{\label{subsec:semisphere}Sound field separation in a room}
In practical implementations of the proposed SFS method, the target 
sources may be placed on a surface. In case the surface is rigid or 
highly reflective, we can measure the sound field using a semi-spherical 
array around the target sources, and duplicate the measurements with 
respect to the surface to use the full spherical harmonics 
expansion.~\cite{Melon2010-cn,Braikia2013-ny}
In this section, we simulate such a case. 

We have a room of size $(4,5,3)$ m as shown in Fig.~8. 
The reflection coefficients of all the walls, floor, and ceiling are 0.99.
One corner of the room is located at $D=(-1.8, -1.5, 0)$ m with respect to 
$O$, a point on the floor. Based on the point $O$, we set up a Cartesian 
and a spherical coordinate system. There is a target point source at the point 
$O$ inside the semi-sphere $\mathbb{S}_2$ of radius $R=0.5$. The length of 
impulse response functions $g_0^{\mu}(t)$, $g_1^{\mu}(t)$, $g_2^{\mu}(t)$ 
is $T_n=\lceil{2R/cf_{\mathrm{s}}}\rceil+1=141$ taps. There is another interfering 
point source at $(0.7,0.8,0.7)$ m outside of the sphere $\mathbb{S}_2$. 
These two sources produces band-limited random noises, which are 
generated by passing unit-variance Gauss white noises through a 64-tap Bandpass 
Butterworth filter with frequency range [100, 300] Hz.

We measure the sound pressure and the radial particle velocity on the semi-sphere 
$\mathbb{S}_2$ at four sensors, whose locations are determined according to the 
first-order Gauss sampling scheme.~\cite{Rafaely2015-yh} We simulate the impulse 
responses, including the radial particle velocity response, between each source 
and sensor using the image source method.~\cite{allen1979image} 
We take the first 2744 image sources into consideration, and  the impulse 
responses are truncated to 8192 taps long under the sampling frequency of 48 kHz.
The pressure and the radial partial velocity on the semi-sphere $\mathbb{S}_2$ are 
duplicated with respect to the floor to use the full spherical harmonic expansions.~\cite{Melon2010-cn,Braikia2013-ny} We estimate the outgoing field 
coefficients up to 0-th order, and reconstruct the outgoing field on the sphere 
over a period of 10 ms according to~\eqref{eq:t_out}. We depict the amplitudes 
of the outgoing field $p^{\mathrm{o}}(t,R,\Theta_{1})$, the total field  $p(t,R,\Theta_{1})$, 
the separated outgoing field $\hat{p}^{\mathrm{o}}(t,R,\Theta_{1})$, and the 
outgoing field separation error $p^{\mathrm{e}}(t,R,\Theta_{1})=p^{\mathrm{o}}(t,R,\Theta_{1})-\hat{p}^{\mathrm{o}}(t,R,\Theta_{1})$,
at the first sensor in Fig.~9. The outgoing field is 
produced by convolving  the target source outputs and corresponding the half-space Green function.~\cite{Melon2010-cn,Braikia2013-ny} The total filed  is the summation 
of the outgoing and incoming fields. 

\begin{figure}[t]
\centering
\includegraphics[width=6cm]{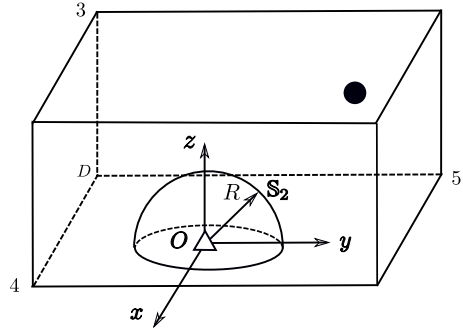}
\caption{Sound field separation in a room: A target source $\triangle$ is 
placed on at the point $O$ inside a semi-sphere $\mathbb{S}_2$ of radius $R=0.5$ m, 
and another point source $\bullet$ is placed at $(0.7,0.8,0.7)$ m with respective 
to the point $O$.}
\label{fig:semisphere}
\end{figure}
\begin{figure}[t]
\centering
\includegraphics[width=9cm]{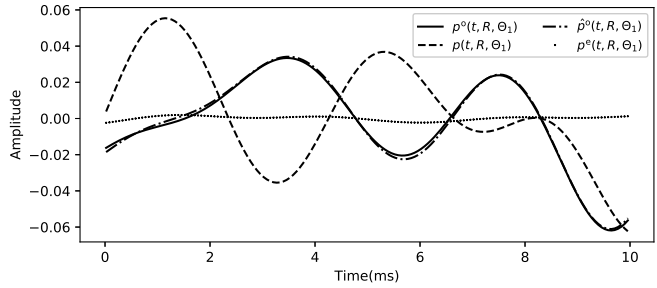}
\caption{Sound field separation at a point: The amplitudes of 
the outgoing field $p^{\mathrm{o}}(t,R,\Theta_{1})$, the total field $p(t,R,\Theta_{1})$, 
the separated outgoing field $\hat{p}^{\mathrm{o}}(t,R,\Theta_{1})$, and the outgoing 
field separation error $p^{\mathrm{e}}(t,R,\Theta_{1})$, at the first sensor 
over 10 ms.} 
\label{fig:semi}
\end{figure}
As shown in Fig.~9, in this case, the proposed method is still able to accurately estimate the outgoing field. The 
normalized separation error at the first sensor is  
\begin{IEEEeqnarray}{rCl}
	\xi_{1}=10\mathrm{log}10\frac{\sum_{t=0}^{10\; \mathrm{ms}}||p^{\mathrm{e}}(t,R,\Theta_{1})||^2}
	{\sum_{t=0}^{10\;\mathrm{ms}}||p^{\mathrm{o}}(t,R,\Theta_{1})||^2}=-31\; \mathrm{dB}.
\end{IEEEeqnarray}
The proposed method also recovers the outgoing field 
over the whole semi-sphere. The results are similar to Fig.~4, 
thus are not shown for brevity.

\section{Conclusion}
In this paper, we developed a time-domain SFS method that can separate 
non-stationary sound fields over a sphere. We decompose the sound field 
and the radial particle velocity measure on the sphere into  spherical 
harmonic coefficients, and recover the outgoing/incoming field 
based on the time-domain relationship between the coefficients
and the derived separation filters.  
The simulations demonstrated that the proposed method
can  separate non-stationary sound fields  in both free field and room 
environments. Further, the method is able to accurately recover the outgoing 
field over a long period of time.  
A future extension of the proposed method is to take the scattering from 
the target source surface into consideration.~\cite{Langrenne2007-hz,
Langrenne2009-mf,Bi2012-aj}
The experimental validation of the proposed  method will be 
our future work.

\begin{acknowledgments}
This work is sponsored by the Australian Research Council (ARC) discovery projects funding 
schemes with project number DP180102375. Fei Ma is supported by the China Scholarship 
Council - Australian National University Joint Funding Program. 
\end{acknowledgments}

\appendix
\section{Proof of theorem 1}
\begin{proof}
	In this section,	we derive the expression of the time-domain outgoing field coefficients $a_{\mu\nu}^{\mathrm{o}}(t,R)$.
	
	Substitutions of $\tau_R=R/c$  and~\cite{Abramowitz1965-oq}
	\begin{IEEEeqnarray}{rCl}
		j_{\mu}^{\prime}(x)= \frac{\mu}{x} j_{\mu}(x)- j_{\mu+1}(x),	\nonumber
	\end{IEEEeqnarray}
	in~\eqref{eq:f_out_coeff} result in   
	\begin{IEEEeqnarray}{rCl}
		\label{eq:f_out_coeff2}
		{A}_{\mu\nu}^{\mathrm{o}}(\omega,R)
		&=&-\mu(i\omega\tau_R)j_{\mu}(\omega{\tau_R})h_{\mu}(\omega{\tau_R})\times{}A_{\mu\nu}(\omega,R)
		\nonumber\\
		&&+\tau_R(\omega{\tau_R})j_{\mu+1}(\omega{\tau_R})h_{\mu}(\omega{\tau_R})
		\times(i\omega)A_{\mu\nu}(\omega,R)	\nonumber\\ 
		&&									
		{-\rho_0{c}}\tau_R(i\omega{\tau_R})j_{\mu}(\omega{\tau_R})h_{\mu} 
		(\omega{\tau_R})\times 
		(i\omega){B}_{\mu\nu}(\omega,R)	
	\end{IEEEeqnarray}
	We define the time-domain outgoing field coefficients  $a_{\mu\nu}^{\mathrm{o}}(t,R)$ as
	\begin{IEEEeqnarray}{rCl}
		a_{\mu\nu}^{\mathrm{o}}(t,R) 
		&\equiv&
		\mathcal{F}^{-1}\{{A}_{\mu\nu}^{\mathrm{o}}(\omega,R)\}	\nonumber	\\
		&=&{g}_0^{\mu}(t) \ast{}\lambda_{\mu\nu}(t)+
		{g}_1^{\mu}(t) \ast{}\chi_{\mu\nu}(t)					
		+
		\rho_0 c\,
		{g}_2^{\mu}(t) \ast{}\eta_{\mu\nu}(t),					
		\label{eq:t_out_coeff}		
	\end{IEEEeqnarray}
	where $\mathcal{F}^{-1}$ denotes the inverse Fourier transform.~\cite{Oppenheim1997-tu} 
	The terms in the second line of~\eqref{eq:t_out_coeff} are defined 
	based on \eqref{eq:f_out_coeff2} as follows
	\begin{IEEEeqnarray}{rCl}
		\lambda_{\mu\nu}(t)&\equiv&\mathcal{F}^{-1}\Big\{{A}_{\mu\nu}(\omega,R)\Big\},  \\
		\chi_{\mu\nu}(t)&\equiv&\mathcal{F}^{-1}\Big\{(i\omega){A}_{\mu\nu}(\omega,R)\Big\},  \\
		\eta_{\mu\nu}(t)&\equiv& \mathcal{F}^{-1}\Big\{(i\omega){ B}_{\mu\nu}(\omega,R)\Big\},\\
		{g}_0^{\mu}(t)
		&\equiv&
		\mathcal{F}^{-1}
		\big\{
		-\mu(i\omega\tau_R)j_{\mu}(\omega{\tau_R})h_{\mu}(\omega{\tau_R})
		\big\},					\label{eq:g0uexp}						\\
		{g}_1^{\mu}(t)
		&\equiv&
		\mathcal{F}^{-1}
		\big\{
		\tau_R(\omega{\tau_R})j_{\mu+1}(\omega{\tau_R})h_{\mu}(\omega{\tau_R})
		\big\},							\label{eq:g1uexp}						\\
		{g}_2^{\mu}(t)
		&\equiv&
		\mathcal{F}^{-1}\big\{-\tau_R(i\omega{}\tau_R){j}_{\mu}(\omega{}\tau_R) h_{\mu} (\omega{}\tau_R)\big\}. 		
		\label{eq:g2uexp}	
		\IEEEeqnarraynumspace
	\end{IEEEeqnarray}
	We derive expressions for these six terms in the following.

	First, based on properties of the Fourier transform and the spherical harmonic 
	transform,~\cite{Williams1999-gk,Abramowitz1965-oq} we obtain $\lambda_{\mu\nu}(t)$, 
	$\chi_{\mu\nu}(t)$, and $\eta_{\mu\nu}(t)$ by decomposing the the pressure 
	$p(t,R,\Theta)$ and the radial particle velocity $v(t,R,\Theta)$ as follows
	\begin{IEEEeqnarray}{rCl}
		\label{eq:t_aa}
		\lambda_{\mu\nu}(t)
		&=&\mathcal{F}^{-1}\{{A}_{\mu\nu}(\omega,R)\} 							\nonumber \\
		&=&\mathcal{F}^{-1}
		\Big\{\int_{\Theta} P(\omega,R,\Theta)Y_{\mu\nu}(\Theta)d\Theta	\Big\}	\nonumber 							 \\
		&=& \int_{\Theta}p(t,R,\Theta)Y_{\mu\nu}(\Theta)d\Theta	  	,
	\end{IEEEeqnarray}
	\begin{IEEEeqnarray}{rCl}
		\label{eq:t_alpha_R1}
		\chi_{\mu\nu}(t)&=& \mathcal{F}^{-1}\Big\{(i\omega){A}_{\mu\nu}(\omega,R)\Big\}	 \nonumber				\\
		&=&\frac{d}{dt}\Big\{\mathcal{F}^{-1}
		[{A}_{\mu\nu}(\omega,R)]\Big\} 							\nonumber \\
		&=&\frac{d}{dt}\Big\{\mathcal{F}^{-1}
		\Big[\int_{\Theta} P(\omega,R,\Theta)Y_{\mu\nu}(\Theta)d\Theta	\Big]\Big\}	\nonumber 							 \\
		&=&\frac{d}{dt}\Big\{ \int_{\Theta} p(t,R,\Theta)Y_{\mu\nu}(\Theta)d\Theta	 \Big\} 	\nonumber\\
		&=& \int_{\Theta} \frac{dp(t,R,\Theta)}{dt}Y_{\mu\nu}(\Theta)d\Theta,
	\end{IEEEeqnarray}
	\begin{IEEEeqnarray}{rCl}
		\label{eq:t_beta_R1}
		\eta_{\mu\nu}(t) 
		&=& \mathcal{F}^{-1}\Big\{(i\omega){ B}_{\mu\nu}(\omega,R)\Big\}	\nonumber \\
		&=&\frac{d}{dt}\Big\{\mathcal{F}^{-1}[{ B}_{\mu\nu}(\omega,R)] \Big\}	\nonumber \\
		&=&\frac{d}{dt}\Big\{\mathcal{F}^{-1}\Big[\int_{\Theta} V(\omega,R,\Theta)Y_{\mu\nu}(\Theta)d\Theta	\Big] \Big\} \nonumber\\
		&=&\frac{d}{dt}\Big\{\int_{\Theta} v(t,R,\Theta)Y_{\mu\nu}(\Theta)d\Theta	 \Big\}	\nonumber\\
		&=&\int_{\Theta} \frac{dv(t,R,\Theta)}{dt}  
		Y_{\mu\nu}(\Theta)d\Theta.
	\end{IEEEeqnarray}
	
	Then based on the following properties of the spherical Bessel and Hankel 
	function~\cite{Williams1999-gk,Abramowitz1965-oq}
	\begin{IEEEeqnarray}{rCl}
		h_{\mu} (x)&=&e^{-ix}i^{\mu+2}\sum_{\nu=0}^{\mu}\frac{\varphi_{\nu}(\mu)}{(ix)^{\nu+1}}, 	\nonumber\\
		\varphi_{\nu}(\mu)&=&\frac{(\mu+\nu)!}{2^{\nu}\nu!(\mu-\nu)!},			\nonumber 	\\
		j_{\mu}(x)&=&\frac{1}{2}[h_{\mu} (x)^{*}+h_{\mu} (x)], 	   												\nonumber 	\\
		h_{\mu}^{\prime}(x) &=&h_{\mu-1}(x)- \frac{\mu+1}{x}h_{\mu}(x), 										\nonumber	\\
		h_{-\mu-1}(x)&=&i(-1)^{\mu+1}h_{\mu}(x),\quad \mu\geq0,													\nonumber
		\label{eq:sphericalbessel}
	\end{IEEEeqnarray}
	we expand \eqref{eq:g0uexp}, \eqref{eq:g1uexp}, and \eqref{eq:g2uexp} as  
	\begin{IEEEeqnarray}{rCl}
		{g}_0^{\mu}(t) 
		&=&\mathcal{F}^{-1}\Big[-{\mu}(i\tau_R\omega)j_{\mu}(\omega{\tau_R})h_{\mu}(\omega{\tau_R})\Big]
		\nonumber\\
		&=&
		\mathcal{F}^{-1}\Big[\frac{\mu}{2}
		\sum_{\nu=0}^{\mu}\sum_{\varsigma=0}^{\mu}\varphi_{\nu}(\mu)\varphi_{\varsigma}(\mu)
		\frac{(-1)^{\nu}+(-1)^{\mu+1}e^{-2i{\tau_R}\omega}}{(i\tau\omega)^{\nu+\varsigma+1}} 
		\Big], 
		\label{eq:g0w}
	\end{IEEEeqnarray}
	\begin{IEEEeqnarray}{rCl}
		{g}_1^{\mu}(t) 
		&=&\mathcal{F}^{-1}\Big[
		\tau_R(\omega{\tau_R})j_{\mu+1}(\omega{\tau_R})h_{\mu}(\omega{\tau_R})\Big]	\nonumber\\
		&=&
		\mathcal{F}^{-1}\Big[
		\frac{\tau_R}{2}
		\sum_{\nu=0}^{\mu+1}\sum_{\varsigma=0}^{\mu}\varphi_{\nu}(\mu+1)\varphi_{\varsigma}(\mu)
		\frac{(-1)^{\nu}+(-1)^{\mu}e^{-2i{\tau_R}\omega}}{(i\tau_R\omega)^{\nu+\varsigma+1}} 
		\Big],
		\label{eq:g1w}
	\end{IEEEeqnarray}
	\begin{IEEEeqnarray}{rCl}
		{g}_2^{\mu}(t)
		&=&\mathcal{F}^{-1}\Big[-\tau_R(i\omega{}\tau_R){j}_{\mu}(\omega{}\tau_R) h_{\mu} (\omega{}\tau_R)\Big] \nonumber\\
		&=&\mathcal{F}^{-1}\Big[\frac{\tau_R}{2}\sum_{\nu=0}^{\mu} 
		\sum_{\varsigma=0}^{\mu}\varphi_{\nu}(\mu)\varphi_{\varsigma}(\mu)
		\frac{(-1)^{\nu}+ (-1)^{\mu+1}e^{-i2\omega{}\tau_R}}{(i\tau_R\omega)^{\nu+\varsigma+1}}\Big],
		\label{eq:g2w} 
	\end{IEEEeqnarray}
	respectively. 
	Based on the Fourier transform properties,~\cite{Oppenheim1997-tu}~\eqref{eq:g0w},~\eqref{eq:g1w}, 
	and \eqref{eq:g2w} 
	can be further developed as \eqref{eq:g0u}, \eqref{eq:g1u}, and \eqref{eq:g2u},  respectively.
	
	The expressions of the incoming field coefficients $a_{\mu\nu}^{\mathrm{i}}(t,R)$, 
	$g_3^{\mu}(t)$, and $g_4^{\mu}(t)$  can be derived similarly, and are not shown for brevity.  
	
\end{proof}

\end{document}